\newcommand{\psfile}[3][]{ 
  \begin{center}
    \setlength{\epsfxsize}{#3\linewidth}\leavevmode
    \def\noOpt{}\def\testit{#1}\ifx\testit\noOpt%
      \epsfbox{#2}%
    \else%
      \epsfbox[#1]{#2}%
    \fi
  \end{center} 
}
\newcommand{\psfiletwoBB}[5]{ 
  \begin{minipage}{\linewidth}
    \parbox[b]{.49\linewidth}{%
      \begin{center}
        \setlength{\epsfxsize}{#5\linewidth}\leavevmode\epsfbox[#1]{#2}
      \end{center}
    }
    \hfill
    \parbox[b]{.49\linewidth}{%
      \begin{center}
        \setlength{\epsfxsize}{#5\linewidth}\leavevmode\epsfbox[#3]{#4}
      \end{center}
    }
  \end{minipage}
}

\documentstyle[aps,prl,preprint,floats,epsf]{revtex}

\newcommand{\goto}{\rightarrow}
\newcommand{\ra}{\rightarrow}
\newcommand{\calB}{\mbox{${\cal B}$}}
\newcommand{\calL}{\mbox{${\cal L}$}}
\newcommand{\calP}{\mbox{${\cal P}$}}
\newcommand{\vbeta}{\mbox{$\vec\beta$}}
\newcommand{\vgamma}{\mbox{$\vec\gamma$}}
\newcommand{\etapr}{\mbox{$\eta^\prime$}}
\newcommand{\etaK}{\mbox{$B\ra\eta K$}}
\newcommand{\etaKst}{\mbox{$B\ra\eta K^*$}}

\newcommand{\BetaKstp}{\mbox{$\calB(B^+\ra\eta K^{*+})$}}
\newcommand{\retaKstp}{\mbox{$27.3^{+9.6}_{-8.2}\pm 5.0$}}
\newcommand{\RetaKstp}{\mbox{$(\retaKstp)\times 10^{-6}$}}
\newcommand{\etaKstz}{\mbox{$B^0\ra\eta K^{*0}$}}
\newcommand{\BetaKstz}{\mbox{$\calB(B^0\ra\eta K^{*0})$}}
\newcommand{\retaKstz}{\mbox{$13.8^{+5.5}_{-4.4}\pm 1.7$}}
\newcommand{\RetaKstz}{\mbox{$(\retaKstz)\times 10^{-6}$}}
\newcommand{\etapK}{\mbox{$B\ra\eta^\prime K$}}

\newcommand{\BetapKp}{\mbox{$\calB(B^+\ra\eta^\prime K^+)$}}
\newcommand{\retapKp}{\mbox{$80^{+10}_{-9}\pm8$}}
\newcommand{\RetapKp}{\mbox{$(\retapKp)\times 10^{-6}$}}
\newcommand{\etapKz}{\mbox{$B^0\ra\eta^\prime K^0$}}
\newcommand{\BetapKz}{\mbox{$\calB(B^0\ra\eta^\prime K^0)$}}
\newcommand{\retapKz}{\mbox{$88^{+18}_{-16}\pm9$}}
\newcommand{\RetapKz}{\mbox{$(\retapKz)\times 10^{-6}$}}

\newcommand{\DE}{\mbox{$\Delta E$}}
\newcommand{\mb}{\mbox{$M$}}
\newcommand{\xf}{\mbox{${\cal F}$}}
\newcommand{\hel}{\mbox{${\cal H}$}}
\newcommand{\piz}{\mbox{$\pi^0$}}
\newcommand{\GeVc}{\mbox{$\text{GeV}/c$}}
\newcommand{\GeVcsq}{\mbox{$\text{GeV}/c^2$}}     

\newcommand{\MeVc}{\mbox{$\text{MeV}/c$}}
\newcommand{\MeVcsq}{\mbox{$\text{MeV}/c^2$}}

\begin{document}

\preprint{\tighten\vbox{\hbox{\hfil CLEO CONF 99-12}
}}

\title{
Two-body \boldmath{$B$} Meson Decays to \boldmath{$\eta$} and
\boldmath{$\eta^\prime$}--Observation of \boldmath{$B\ra\eta K^*$}
}  

\author{CLEO Collaboration}
\date{\today}

\maketitle
\tighten

\begin{abstract} 

In a sample of 19 million produced $B$ mesons we have observed decays
\etaKst, and improved our previous measurements of \etapK.  The 
branching fractions we measure for these decay modes are $\BetaKstp =
\RetaKstp$, $\BetaKstz = \RetaKstz$, $\BetapKp = \RetapKp$ and  
$\BetapKz = \RetapKz$.  We have searched with comparable sensitivity for
related nonstrange decays, and report upper limits for these
rates.  All quoted results are preliminary.

\end{abstract}
\newpage

{
\renewcommand{\thefootnote}{\fnsymbol{footnote}}

\begin{center}
S.~J.~Richichi,$^{1}$ H.~Severini,$^{1}$ P.~Skubic,$^{1}$
A.~Undrus,$^{1}$
M.~Bishai,$^{2}$ S.~Chen,$^{2}$ J.~Fast,$^{2}$
J.~W.~Hinson,$^{2}$ J.~Lee,$^{2}$ N.~Menon,$^{2}$
% ...
\end{center}
 
\small
\begin{center}
$^{1}${University of Oklahoma, Norman, Oklahoma 73019}\\
$^{2}${Purdue University, West Lafayette, Indiana 47907}\\
% ...
\end{center}

\setcounter{footnote}{0}
}
\newpage

\begin{figure}[htbp]
\psfile[105 380 540 695]{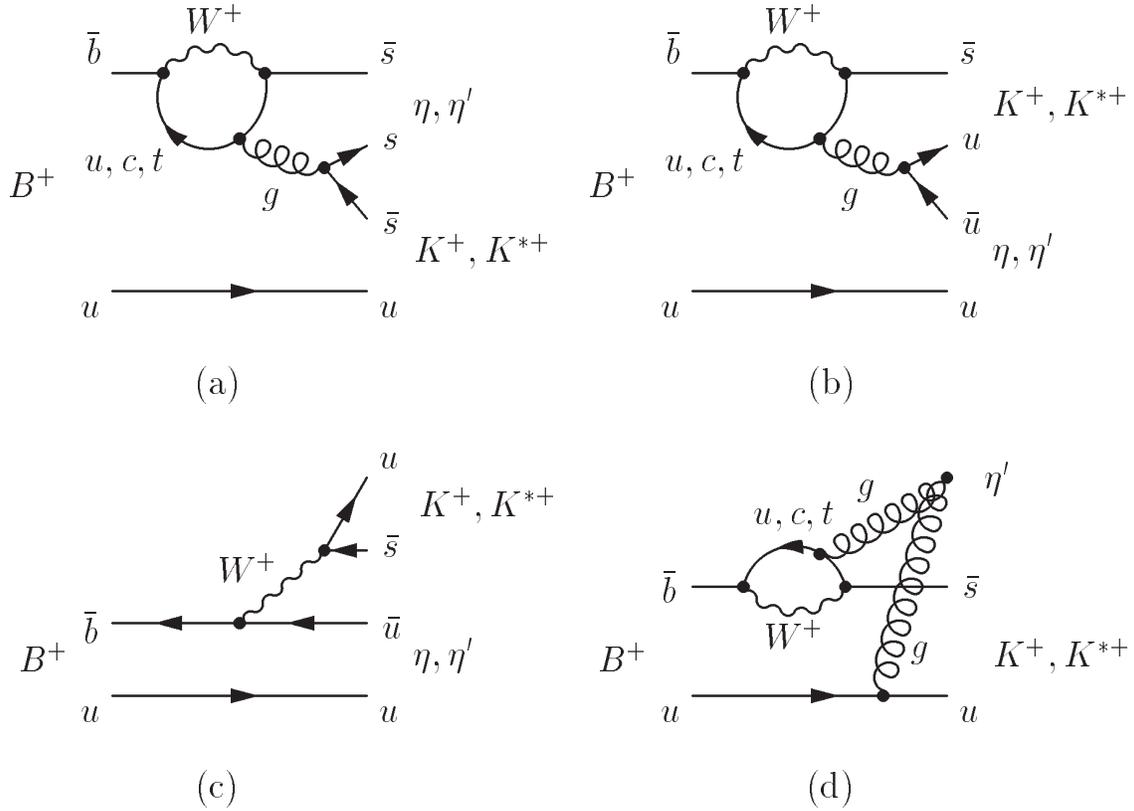}{0.9}
 \caption{\label{fig:diagrams}%
Feynman diagrams describing the representative decays
$B^+\ra\eta^{(\prime)}K^{(*)+}$:  (a, b) internal penguins; (c)
external tree; (d) flavor-singlet penguin.
 }  
\end{figure}

The dominant decay modes of $B$ mesons involve the $\bar b\ra \bar c$
quark transition with coupling to a $W^+$ boson.  For many of these
modes the decay amplitude may be described by a tree diagram in which
the light quark (spectator) is bound in both the initial $B$ meson and
final charmed hadron via soft gluon exchange.  With recent improvements
in experimental sensitivity, less favored modes are becoming accessible.
These include: $b\ra u$ tree diagram transitions that are suppressed
by the small Cabibbo-Kobayashi-Maskawa \cite{ckm} (CKM) matrix element
$V_{ub}$, such as $B\ra\pi\ell\nu$
\cite{CLEObpilnu}; effective flavor changing neutral current (FCNC)
decays $b\ra s$ described by loop diagrams, such as the ``electromagnetic
penguin'' $B\ra K^*\gamma$ \cite{CLEObkstg}; and
decays to charmless hadrons such as $B\ra K\pi$
\cite{CLEObkpi,bigrare,CLEObkpiNew}.  The hadronic decays 
may be classified according to contributions to the amplitude from the
several tree and penguin diagrams shown in Fig.\ \ref{fig:diagrams}
\cite{thyPred,aliGreub}.  Some of these charmless hadronic decays offer
prospects for the observation of $CP$ violation, while others
facilitate the quantitative understanding of the amplitudes that are
essential to the interpretation of future $CP$ measurements.  For
example, the decays \etaK\ and \etapK, with $B\ra K\pi$, have been
examined in this context \cite{kps,etaCP}.

In this paper we present results of several new experimental searches
for $B$ meson decays to two-body final states containing $\eta$ and
$\eta^\prime$ mesons.  These $I=0$ mesons are mixtures of flavor-SU(3)
octet and singlet states, the latter being of particular interest
because of its allowed formation through a pure (two or more) gluon
intermediate state (Fig.\ \ref{fig:diagrams} (d)).

The data were accumulated at the Cornell Electron-positron Storage Ring
(CESR).  The integrated luminosity was 9.13 $\text{fb}^{-1}$ for the
reaction $e^+e^-\ra\Upsilon(4S)\ra B\bar B$ (center-of-mass energy
$E_{\rm cm}=10.58$ GeV). This luminosity corresponds to the production
of $9.6\times10^6$ charged and an approximately equal number of neutral
$B$ mesons.  In addition we recorded 4.35 ${\rm fb}^{-1}$ of data with
$E_{\rm cm}$ below the threshold for $B\bar B$ production to measure
continuum processes.  These constitute the complete data sample from the
CLEO II and CLEO II.V experiments, and the measurements we report here
succeed our earlier ones \cite{cleoEtaPRL,cleoVanc}\ from subsets of these
data.

The CLEO II detector\cite{CLEOdet}\ emphasizes precision charged
particle tracking, with specific ionization ($dE/dx$) measurement, and
high resolution electromagnetic calorimetry based on CsI(Tl).
Scintillators between the tracking chambers and calorimeter provide 
time-of-flight information which we use in conjuction with $dE/dx$ for
particle identification (PID).  
The CLEO II.V configuration differs in two respects: the addition of a
silicon vertex detector; and replacement of the 50:50 argon:ethane gas
in the main drift chamber with a 60:40 helium:propane mixture.  From the
raw data we reconstruct charged pions and kaons, photons (from $\pi^0$,
$\eta$, and $\eta^\prime$ decays), and $\pi^+\pi^-$ pairs that intersect
at a vertex displaced from the
collision point (``vees'', from $K^0_s\ra\pi^+\pi^-$).  Candidate $B$
decay tracks must meet specifications on the number of drift chamber
measurements, goodness of fit, and consistency with an origin at the
primary or particular secondary vertex.  Candidate photons must be
isolated calorimeter clusters with a photon-like spatial distribution
and energy deposition exceeding 30 MeV.  We exclude photon pairs from
extremely asymmetric $\eta$ decays,
requiring $\left|\cos\theta^*\right|<0.97$, where
$\theta^*$ is the meson center of mass decay angle relative to its
flight direction.  This cut rejects soft photon backgrounds and is
tightened to 0.90 for $K^*/\rho$ channels to veto $B\ra K^*\gamma$
background.  We reject charged tracks and photon pairs having
momentum less than $100\ \MeVc$.  The photon from candidate
$\etapr\ra\rho\gamma$ decays is generally required to have an energy
greater than 200 MeV, though this requirement is relaxed to 100 MeV for
channels with a $K_S$, which have low background.

We fit photon pairs and vees kinematically to the appropriate combined
mass hypothesis to obtain the meson momentum vectors.  Resolutions on
the reconstructed masses prior to the constraint are about 5 -- 10
$\MeVcsq$ (momentum dependent)
for $\pi^0\ra\gamma\gamma$, $12\ \MeVcsq$ for $\eta\ra\gamma\gamma$,
and $3\ \MeVcsq$ for $K^0_s\ra\pi^+\pi^-$.  Information about expected
signal distributions with the detector response comes from a detailed
GEANT-based simulation of the CLEO detector \cite{GEANT}\ that
reproduces the 
resolutions and efficiencies of data in a variety of benchmark
processes.  
In particular, we have established the momentum and $dE/dx$
resolutions in studies of $D^0\ra K^-\pi^+$ decays in a
momentum range near $2.6\ \GeVc$.

Since the $B$ mesons are formed nearly at rest, while the $B$ daughters
we observe are relatively light, the latter have momenta close to half
of the beam energy ($2.6\ \GeVc$).  For this reason the final states are
well separated from those involving heavier daughters, i.e., the
dominant $b\ra c$ decays.  The principal signatures for the selected
decay modes are consistency of the resonance decay invariant masses with
the known masses and widths of those resonances, and consistency of the
total final state with the $B$ meson mass and energy.  Because the beam
energy $E_b$ is better known than the reconstructed $B$ meson energy
$E_B$, we substitute the former in the $B$ mass calculation: $\mb
\equiv \sqrt{E_b^2-{\bf p}_B^2}$, with ${\bf p}_B$ the reconstructed $B$
momentum.  We define also the variable $\Delta E\equiv E_B-E_b$.  The
measurement resolution on \mb\ is 2.5 -- 3 $\MeVcsq$, and on \DE\ it is
25-50 MeV, depending on the apportionment of the energy among charged
tracks and photons for each mode.

For vector-pseudoscalar decays of the $B$ and the $\rho\gamma$ decay of the
$\eta^\prime$ we gain further discrimination from the
helicity variable \hel\ (cosine of the vector meson's rest frame
two-body decay angle with respect to its flight direction), which
reflects the spin alignment in the decay. 
The decay rate is proportional to $\hel^2$ when the vector meson decays
into two spinless daughters, and to $1-\hel^2$ for $\etapr\ra\rho\gamma$.
For modes in which one daughter is a single
charged track
we
achieve statistical discrimination between kaons and pions by $dE/dx$.
With $S_K$ and $S_\pi$ defined as the
deviations from nominal energy loss for the indicated particle
hypotheses measured in standard deviations, the separation $S_K-S_\pi$
is about 1.7 (2.0) at $2.6\ \GeVc$ for the CLEO II (II.V) samples. 

The main backgrounds arise from continuum quark production
$e^+e^-\ra q\bar q$.  We discriminate against these jet-like events
with several measures of the energy flow pattern.  One is the angle
$\theta_{BB}$ between the thrust axis (axis of maximum energy projection
magnitude) of the candidate $B$ and that of the rest of the event.  For
a fake $B$ candidate selected from particles belonging to a $q\bar q$
event, those particles tend to align with the rest of the event, whereas
the true $B$ decays have a thrust axis that is largely uncorrelated with the
tracks and showers from the decay of the partner $B$.  We reject events
with $\left|\cos\theta_{BB}\right|>0.9$.  In addition we use a
multivariate discriminant \xf\ incorporating the energy
deposition in nine cones concentric with the event thrust axis, and the
angles of the thrust axis and ${\bf p}_B$ with respect to the $e^+e^-$
beam direction \cite{bigrare}.  We have checked the
backgrounds from the favored $B$ decay modes by simulation, finding their
contributions to the modes in this study to be generally quite small.
Where appropriate we include this component in the fits described
below.  

\def\sgline{\noalign{\vskip 0.15truecm\hrule\vskip 0.15truecm}}
\def\piz{\pi^0}
\def\calB{\mbox{${\cal B}$}}
\newcommand{\etaprk}{\mbox{$\etapr K$}}
\newcommand{\etaprkp}{\mbox{$\etapr K^+$}}
\newcommand{\etaprkpd}{\mbox{$\etapr_{\eta\pi\pi}K^+$}}
\newcommand{\etaprkprg}{\mbox{$\etapr_{\rho\gamma}K^+$}}
\newcommand{\etaprkpfv}{\mbox{$\etapr_{5\pi}K^+$}}
\newcommand{\etaprkz}{\mbox{$\etapr K^0$}}
\newcommand{\etaprkzd}{\mbox{$\etapr_{\eta\pi\pi} K^0$}}
\newcommand{\etaprkzrg}{\mbox{$\etapr_{\rho\gamma} K^0$}}
\newcommand{\etaprpi}{\mbox{$\etapr\pi^+$}}
\newcommand{\etaprpid}{\mbox{$\etapr_{\eta\pi\pi}\pi^+$}}
\newcommand{\etaprpirg}{\mbox{$\etapr_{\rho\gamma}\pi^+$}}
\newcommand{\etaprpifv}{\mbox{$\etapr_{5\pi}\pi^+$}}
\newcommand{\etaprh}{\mbox{$\etapr h^+$}}
\newcommand{\etaprpiz}{\mbox{$\etapr\piz$}}
\newcommand{\etaprpizepp}{\mbox{$\etapr_{\eta\pi\pi}\piz$}}
\newcommand{\etaprpizrg}{\mbox{$\etapr_{\rho\gamma}\piz$}}
\newcommand{\etaprkstz}{\mbox{$\etapr K^{*0}$}}
\newcommand{\etaprkstzd}{\mbox{$\etapr_{\eta\pi\pi} K^{*0}$}}
\newcommand{\etaprkstp}{\mbox{$\etapr K^{*+}$}}
\newcommand{\etaprkstpd}{\mbox{$\etapr_{\eta\pi\pi} K^{*+}_{K^+\piz}$}}
\newcommand{\etaprkstpkz}{\mbox{$\etapr_{\eta\pi\pi} K^{*+}_{K^0\pi^+}$}}
\newcommand{\etaprrhoz}{\mbox{$\etapr\rho^0$}}
\newcommand{\etaprrhozd}{\mbox{$\etapr_{\eta\pi\pi}\rho^0$}}
\newcommand{\etaprrhop}{\mbox{$\etapr\rho^+$}}
\newcommand{\etaprrhopd}{\mbox{$\etapr_{\eta\pi\pi}\rho^+$}}
\newcommand{\etapreta}{\mbox{$\etapr\eta$}}
\newcommand{\etapretagg}{\mbox{$\etapr_{\eta\pi\pi}\eta_{\gaga}$}}
\newcommand{\etapretathrp}{\mbox{$\etapr_{\eta\pi\pi}\eta_{3\pi}$}}
\newcommand{\etapretarg}{\mbox{$\etapr_{\rho\gamma}\eta_{\gaga}$}}
\newcommand{\etapretargtp}{\mbox{$\etapr_{\rho\gamma}\eta_{3\pi}$}}
\newcommand{\etapretapr}{\mbox{$\etapr\etapr$}}
\newcommand{\etapretaprd}{\mbox{$\etapr_{\eta\pi\pi}\etapr_{\eta\pi\pi}$}}
\newcommand{\etapretaprrg}{\mbox{$\etapr_{\eta\pi\pi}\etapr_{\rho\gamma}$}}
\newcommand{\Betaprkp}{\mbox{$B^+\ra\etapr K^+$}}
\newcommand{\Betaprkz}{\mbox{$B^0\ra\etapr K^0$}}
\newcommand{\Betaprpi}{\mbox{$B^+\ra\etapr\pi^+$}}
\newcommand{\Betaprpiz}{\mbox{$B^0\ra\etapr\piz$}}
\newcommand{\Betaprkstz}{\mbox{$B^0\ra\etapr K^{*0}$}}
\newcommand{\Betaprkstp}{\mbox{$B^+\ra\etapr K^{*+}$}}
\newcommand{\Betaprrhoz}{\mbox{$B\ra\etapr\rho^0$}}
\newcommand{\Betaprrhop}{\mbox{$B^+\ra\etapr\rho^+$}}
\newcommand{\Betapreta}{\mbox{$B^0\ra\etapr\eta$}}
\newcommand{\Betapretapr}{\mbox{$B\ra\etapr\etapr$}}

\newcommand{\etak}{\mbox{$\eta K^+$}}
\newcommand{\etakgg}{\mbox{$\eta_{\gaga} K^+$}}
\newcommand{\etakthrp}{\mbox{$\eta_{3\pi} K^+$}}
\newcommand{\etapi}{\mbox{$\eta\pi^+$}}
\newcommand{\etapigg}{\mbox{$\eta_{\gaga}\pi^+$}}
\newcommand{\etapithrp}{\mbox{$\eta_{3\pi}\pi^+$}}
\newcommand{\etapiz}{\mbox{$\eta\piz$}}
\newcommand{\etapizgg}{\mbox{$\eta_{\gaga}\piz$}}
\newcommand{\etapizthrp}{\mbox{$\eta_{3\pi}\piz$}}
\newcommand{\etakz}{\mbox{$\eta K^0$}}
\newcommand{\etakzgg}{\mbox{$\eta_{\gaga} K^0$}}
\newcommand{\etakzthrp}{\mbox{$\eta_{3\pi} K^0$}}
\newcommand{\etaeta}{\mbox{$\eta\eta$}}
\newcommand{\etaetagg}{\mbox{$\eta_{\gaga}\eta_{\gaga}$}}
\newcommand{\etaetathrp}{\mbox{$\eta_{\gaga}\eta_{3\pi}$}}
\newcommand{\etaetasixp}{\mbox{$\eta_{3\pi}\eta_{3\pi}$}}
\newcommand{\etakstz}{\mbox{$\eta K^{*0}$}}
\newcommand{\etakstzgg}{\mbox{$\eta_{\gaga} K^{*0}$}}
\newcommand{\etakstzthrp}{\mbox{$\eta_{3\pi} K^{*0}$}}
\newcommand{\etakstp}{\mbox{$\eta K^{*+}$}}
\newcommand{\etakstpgg}{\mbox{$\eta_{\gaga} K^{*+}_{K^+\piz}$}}
\newcommand{\etakstpthrp}{\mbox{$\eta_{3\pi} K^{*+}_{K^+\piz}$}}
\newcommand{\etakstpggkz}{\mbox{$\eta_{\gaga} K^{*+}_{K^0\pi^+}$}}
\newcommand{\etakstpthrpkz}{\mbox{$\eta_{3\pi} K^{*+}_{K^0\pi^+}$}}
\newcommand{\etarhoz}{\mbox{$\eta \rho^0$}}
\newcommand{\etarhozgg}{\mbox{$\eta_{\gaga} \rho^0$}}
\newcommand{\etarhozthrp}{\mbox{$\eta_{3\pi} \rho^0$}}
\newcommand{\etarhop}{\mbox{$\eta \rho^+$}}
\newcommand{\etarhopgg}{\mbox{$\eta_{\gaga} \rho^+$}}
\newcommand{\etarhopthrp}{\mbox{$\eta_{3\pi} \rho^+$}}
\newcommand{\Betak}{\mbox{$B^+\ra\eta K^+$}}
\newcommand{\Betapi}{\mbox{$B^+\ra\eta\pi^+$}}
\newcommand{\Betapiz}{\mbox{$B^0\ra\eta\piz$}}
\newcommand{\Betakz}{\mbox{$B^0\ra\eta K^0$}}
\newcommand{\Betaeta}{\mbox{$B^0\ra\eta\eta$}}
\newcommand{\Betakstz}{\mbox{$B^0\ra\eta K^{*0}$}}
\newcommand{\Betakstp}{\mbox{$B^+\ra\eta K^{*+}$}}
\newcommand{\Betarhoz}{\mbox{$B^0\ra\eta \rho^0$}}
\newcommand{\Betarhop}{\mbox{$B^+\ra\eta \rho^+$}}

\newcommand{\gaga}{{\gamma\gamma}}
\newcommand{\etagg}{\mbox{$\eta\ra\gaga$}}
\newcommand{\etathrpi}{\mbox{$\eta\ra\pi^+\pi^-\pi^0$}}
\newcommand{\etaprd}{\mbox{$\etapr\ra\eta\pi^+\pi^-$}}
\newcommand{\etaprrg}{\mbox{$\etapr\ra\rho^0\gamma$}}
\newcommand{\ksppd}{\mbox{$K^{0}\rightarrow K_S\rightarrow\pi^+\pi^-$}}
\newcommand{\kstzd}{\mbox{$K^{*0}\ra\K^+\pi^-$}}
\newcommand{\kstpd}{\mbox{$K^{*+}\ra\K^+\piz$}}
\newcommand{\kstpkz}{\mbox{$K^{*+}\ra\K^0\pi^+$}}

\begin{table}[htbp]
\vbox{
\caption{Measurement results.  The first column lists the final states,
with secondary decay modes as subscripts.  The vector modes with
secondary decays  to $h^+\pi^0$ or $h^+\pi^-$ are further distinguished
with a subscript $(-)$ or $(+)$ depending on the value of \hel\ (see text).  
The remaining columns give event yield from the fit, 
reconstruction efficiency $\epsilon$, total efficiency with secondary
branching fractions ${\cal B}_s$, significance, and the resulting $B$
decay branching fraction ${\cal B}$ with upper limit.} 
\def\notext{\noalign{\vskip 0.35truecm}}
\begin{center}
\begin{tabular}{lcrrccc}
Final state & Fit events & $\epsilon$(\%) & $\epsilon\calB_s$(\%) &
              Signif. ($\sigma$) & \calB($10^{-6})$ & 90\%\ UL($10^{-6})$ \cr
\sgline
\etaprkpd     & $39.6^{+7.0}_{-6.4}$ & 27 & 4.7 & 13.4 & 
                $88^{+16}_{-14}$ & $-$ \cr
\etaprkprg  & $61^{+11}_{-10}$ & 29 & 8.7 & 10.1 & 
		$72^{+13}_{-12}$ & $-$ \cr
\etaprkzd      & $9.2^{+3.6}_{-2.9}$ & 24 & 1.4 &  7.7 & 
		$67^{+26}_{-21}$ & $-$ \cr
\etaprkzrg    & $29.6^{+7.0}_{-6.2}$ & 28 & 2.9 &  8.9 & 
               $105^{+25}_{-22}$ & $-$ \cr
\notext
\etaprpid      & 
$0.0^{+2.2}_{-0.0}$ & 28 & 4.7 &  0.0 & $0.0^{+4.9}_{-0.0}$ & 13 \cr
\etaprpirg     & 
$4.4^{+7.2}_{-4.4}$ & 30 & 9.0 &  0.8 & $5.1^{+8.3}_{-5.1}$ & 21 \cr
\notext
\etakgg   &
$5.9^{+6.0}_{-4.6}$ & 45 & 17.5 & 1.4 & $3.5^{+3.5}_{-2.7}$ & 10 \cr
\etakthrp &
$0.0^{+2.0}_{-0.0}$ & 29 &  6.6 & 0.0 & $0.0^{+3.1}_{-0.0}$ &  9.4 \cr
\etakzgg  &
$0.0^{+2.6}_{-0.0}$ & 38 &  5.1 & 0.0 & $0.0^{+5.2}_{-0.0}$ & 12 \cr
\etakzthrp     &
$0.0^{+0.9}_{-0.0}$ & 25 &  1.9 & 0.0 & $0.0^{+5.0}_{-0.0}$ & 23 \cr
\notext
\etapigg  &
$5.7^{+5.7}_{-4.6}$ & 46 & 18.2 & 1.3 & $3.2^{+3.3}_{-2.6}$ &  9.2 \cr
\etapithrp &
 $0.0^{+1.1}_{-0.0}$ & 30 &  6.8 & 0.0 & $0.0^{+1.7}_{-0.0}$ &  7.0 \cr
\etapizgg &
 $0.0^{+1.0}_{-0.0}$ & 35 & 13.7 & 0.0 & $0.0^{+0.8}_{-0.0}$ &  3.9 \cr
\etapizthrp    &
 $0.0^{+1.4}_{-0.0}$ & 20 &  4.6 & 0.0 & $0.0^{+3.1}_{-0.0}$ & 11 \cr
\notext
$\etakstpgg_{(-)}$ & 
$9.2^{+4.4}_{-3.5}$ & 12 & 1.6 & 4.0 & $59^{+28}_{-23}$ & 125 \cr
$\etakstpgg_{(+)}$ & 
$0.0^{+1.3}_{-0.0}$ & ~9 & 1.2 & 0.0 & $~0.0^{+12}_{-0}$ & 53 \cr
$\etakstpthrp_{(-)}$ &
$1.1^{+1.7}_{-1.1}$ & ~8 & 0.7 & 1.4 & $17^{+27}_{-17}$ & 96 \cr
$\etakstpthrp_{(+)}$ &
$3.8^{+3.7}_{-2.5}$ & ~6 & 0.5 & 1.9 & $80^{+77}_{-53}$ & 269 \cr
\etakstpggkz &
$3.3^{+3.0}_{-2.1}$ & 24 & 2.2 & 2.1 & $16^{+14}_{-10}$ &  44 \cr
\etakstpthrpkz &
$3.0^{+2.7}_{-1.9}$ & 17 & 0.9 & 2.1 & $34^{+30}_{-21}$ &  98 \cr
\notext
$\etakstzgg_{(-)}$ &
$6.9^{+3.9}_{-3.0}$ & 16 & 4.3 & 4.2 & $16.7^{+9.4}_{-7.0}$ &  36 \cr
$\etakstzgg_{(+)}$ &
$0.0^{+2.1}_{-0.0}$ & 15 & 4.0 & 0.0 & $~0.0^{+5.4}_{-0.0}$ &  17 \cr
$\etakstzthrp_{(-)}$ &
$4.7^{+3.1}_{-2.3}$ & 11 & 1.7 & 3.8 & $28^{+19}_{-14}$ &  68 \cr
$\etakstzthrp_{(+)}$ &
$2.7^{+3.9}_{-2.7}$ & 10 & 1.6 & 1.0 & $18^{+25}_{-18}$ &  69 \cr
\notext
$\etarhopgg_{(-)}$ &
$0.0^{+1.5}_{-0.0}$ & 12 & 4.8 & 0.0 & $~0.0^{+3.2}_{-0.0}$ &  14 \cr
$\etarhopgg_{(+)}$ &
$0.4^{+4.4}_{-0.4}$ & ~9 & 3.6 & 0.1 & $~1^{+12}_{-1}$ &  31 \cr
$\etarhopthrp_{(-)}$ &
$3.3^{+3.2}_{-2.1}$ & ~8 & 1.9 & 2.1 & $18^{+17}_{-12}$ &  58 \cr
$\etarhopthrp_{(+)}$ &
$1.4^{+3.6}_{-1.4}$ & ~6 & 1.4 & 0.6 & $10^{+26}_{-10}$ & 118 \cr
\notext
\etarhozgg & 
$2.0^{+3.2}_{-2.4}$ & 26 & 10.3 & 1.1 & $~2.0^{+3.3}_{-2.0}$ &  10 \cr
\etarhozthrp &
$2.3^{+4.3}_{-2.3}$ & 18 &  4.2 & 0.7 & $~6^{+11}_{-6}$ & 28 \cr
\end{tabular}
\end{center}
\label{individtab}
}
\end{table}

We set the selection criteria for mass, energy, and event shape variables so
as to include sidebands about the expected signal peaks. 
To extract event yields we perform unbinned extended maximum likelihood
fits to these data
of a superposition of expected signal and background distributions:
\begin{equation}\label{eq:lfit}
 \calL(N_S,N_B) = {e^{-(N_S+N_B)}\over N!} \prod_{i=1}^N \left[N_{S}
               \calP_{S}(\vbeta;{\bf x}_i) + 
               N_B \calP_{B}(\vgamma;{\bf x}_i)\right].
\end{equation}
Here $\calP_S$ and $\calP_B$ are the probability distribution functions
(PDFs) for signal and continuum background, respectively.  They are
functions of observables ${\bf x}_i$ for event $i$, and of parameters
\vbeta\ and \vgamma\ (discussed below).  The form of \calL\ reflects the
underlying Poisson statistics obeyed by $N_S$ and $N_B$, the
(non-negative) numbers of signal and continuum background events,
respectively.  At the maximum of \calL, $N_S+N_B$ is equal to the total
number $N$ of input events.  Observables for each event include \mb,
\DE, \xf, and (where applicable) resonance masses and \hel.
For these fits $N$ ranges from $\sim 100$ to a few thousand.

For $B^+$ decays \cite{chgconj}\ that
have a primary daughter charged hadron (generically
$h^+$) that can be either $\pi^+$ or $K^+$ we fit both modes simultaneously,
with \calL\ expanded so that the signal and background yields of both
$\pi^+$ and $K^+$ are fit variables.  In this case the PDFs depend also
on the $dE/dx$ observables $S_\pi$ and $S_K$.  The modes with a
secondary vector decay $(K^*\ {\rm or}\ \rho)\ra h^+(\pi^-\ {\rm
or}\ \pi^0)$ also require special treatment.  For these modes the momentum
spectrum of $h^+$ is bimodal because of the forward-backward peaked
\hel\ distribution.  We select independent
$K^*$ and $\rho$ samples to fit according to the sign of $\hel$.  
Events with $\hel<0$ in our sign convention
have low momentum $h^+$ and are unambiguously separated
by kinematics combined with PID information from $dE/dx$ measurements.  
For
the events with $\hel>0$ the separation is less sharp, so we allow
the yields for both $h^+$ hypotheses to float in the global likelihood
fit and include the PID information.
For $K^{*0}/\rho^0$ the choice of
$K$ or $\pi$ is made on the basis of $dE/dx$ and time-of-flight
information.

The PDFs $\calP_S$ and $\calP_B$ are constructed as products of
functions of the observables ${\bf x}_i$.  The dependences of $\calP_S$ on
masses and energies are Gaussian, double Gaussian, or Breit-Wigner
functions, whose means, widths, etc.\ appear as the parameters
\vbeta\ in Eq.\ \ref{eq:lfit}.  The background PDF $\calP_B$ contains
signal-like peaking 
components in its resonance mass projections, to account for real
resonances in the background, added to smooth components for
combinatoric continuum.  The smooth components are low-order polynomials,
except that for \mb\ we use an empirical shape
\cite{argus}\ that accounts for the phase space limit at $M=E_b$.  
The dependences of both $\calP_S$ and $\calP_B$ on \xf, $S_K$, and
$S_\pi$ are bifurcated Gaussian functions.  We obtain the parameters
\vbeta\ of $\calP_S$ from separate fits to simulated signal, and \vgamma\ of
$\calP_B$ from fits to data in a sideband region of the $\DE$ -- $\mb$
plane.  Where the Monte Carlo estimate of background from $\Upsilon(4S)$
production is non-negligible, we add a term with a free fit variable to
account for this as well.

Results for all of the $B$ decay chains appear
in Table \ref{individtab}.  The row label subscripts denote secondary
decays, including $\eta^\prime\ra \eta\pi^+\pi^-$ with
$\eta\ra\gamma\gamma$ ($\eta\pi\pi$), 
$\eta^\prime\ra \rho\gamma$ ($\rho\gamma$), and $\eta\ra\pi^+\pi^-\pi^0$
($3\pi$).  For each mode the table gives the event yield from the
likelihood fit, reconstruction efficiency, total efficiency,
significance, branching fraction, and (where appropriate) 90\%
confidence level upper limit.  Significance is defined as the number of
standard deviations corresponding to the probability for a fluctuation
from zero to our observed yield.

\begin{table}[htbp]
\caption{Combined branching fractions.  The fourth column gives our
preliminary result, as central value followed by statistical and
systematic error if significantly above zero, otherwise as a 90\%
confidence level upper limit.  We quote estimates from various theoretical
sources for comparison.}
\def\notext{ & & & & &\cr}
\begin{center}
\begin{tabular}{lccccl}
Decay mode & $\calB_{\rm fit}(10^{-6})$ & Signif. ($\sigma$) & $\calB(10^{-6})$ 
                                        & Source & Theory \calB($10^{-6}$)\cr
\sgline
$B^+\goto\etaprkp$    & \retapKp & 16.8 & \retapKp &
                        This Exp. & 7 -- 65 \cite{chau,du}\cr
$B^0\goto\etaprkz$    & \retapKz & 11.7 & \retapKz &
                        This Exp. & 9 -- 59 \cite{chau,du}\cr
$B^+\goto\etaprpi$    & $1.0^{+5.3}_{-1.0}\pm 0.1$ & $0.2$ & $<11$ &
                        This Exp. & 1 -- 23 \cite{chau,du}\cr
$B^0\goto\etaprpiz$   & & & $<11$ & 
                        \cite{cleoEtaPRL} & 0.1 -- 14 \cite{chau,du}\cr
$B^+\goto\etaprkstp$  & & 1.2 & $<87$ & 
                        \cite{cleoDPF} & 0.1 -- 3.7 \cite{chau,du}\cr
$B^0\goto\etaprkstz$  & &1.0 & $<20$ & 
                        \cite{cleoDPF} & 0.1 -- 8.0 \cite{chau,du}\cr
$B^+\goto\etaprrhop$  & & & $<47$ & 
                        \cite{cleoEtaPRL} & 3 -- 24 \cite{chau,du}\cr
$B^0\goto\etaprrhoz$  & & & $<23$ & 
                        \cite{cleoEtaPRL} & 0.1 -- 11 \cite{chau,du}\cr
$B^+\goto\etak$       & $2.2^{+2.6}_{-2.2}$ & 1.0 & $<7.1$ &
                        This Exp. & 0.2 -- 5.0 \cite{chau,du}\cr
$B^0\goto\etakz$      & $0.0^{+3.0}_{-0.0}$ & 0.0 & $<9.5$ &
                        This Exp. & 0.1 -- 3.0 \cite{chau,dean,du}\cr
$B^+\goto\etapi$      & $1.2^{+2.6}_{-1.2}$ & 0.6 & $<6.0$ &
                        This Exp. & 1.9 -- 7.4 \cite{chau,dean,du}\cr
$B^0\goto\etapiz$     & $0.0^{+0.7}_{-0.0}$ & 0.0 & $<3.1$ &
                        This Exp. & 0.2 -- 4.3 \cite{chau,du}\cr
$B^+\goto\etakstp$    & \retaKstp & 4.8 & \retaKstp &
                        This Exp. & 0.2 -- 8.2 \cite{chau,du}\cr
$B^0\goto\etakstz$    & \retaKstz & 5.1 & \retaKstz & 
                        This Exp. & 0.1 -- 8.9 \cite{chau,dean,du}\cr
$B^+\goto\etarhop$    & $4.3^{+4.6}_{-3.4}\pm 0.7$ & 1.3 & $<16$ &
                        This Exp. & 4 -- 17 \cite{chau,dean,du}\cr
$B^0\goto\etarhoz$    & $2.6^{+3.0}_{-2.4}\pm 0.3$ & 1.3 & $<11$ &
                        This Exp. & 0.1 -- 6.5 \cite{chau,dean,du}\cr
\end{tabular}
\end{center}
\label{combtab}
\end{table}

Where we have measured a given $B$ decay mode in more than one secondary
decay channel, we combine the samples by adding the $\chi^2=-2\ln{\cal
L}$ functions of branching fraction and extracting a value with errors
or limit from the combined distribution.  The limit is the value of
${\cal B}$ below which 90\% of the integral of ${\cal L}$ lies.  We
summarize in Table \ref{combtab}\ the results for these measurements,
along with some earlier ones for related decays, and
corresponding theoretical estimates \cite{kps,chau,dean,du}.
We include systematic errors from
uncertainties in the PDFs, i.e., in \vbeta\ and \vgamma, obtained from a
Monte Carlo convolution of the likelihood function with Gaussian
resolution functions for these parameters, including their most
important correlations.  This procedure changes the upper limit by less
than 10\% in most cases.  We also include systematic errors for
reconstruction efficiencies and selection requirements, and quote upper
limits computed with efficiencies one standard deviation below nominal.

We have analyzed each of the decays also without use of the likelihood
fit, employing more restrictive cuts in each of the variables to isolate
the signals.  The results are consistent with those quoted in the
tables, but with larger errors (less restrictive limits) in most cases.

\begin{figure}[htbp]
\psfiletwoBB{60 150 530 610}{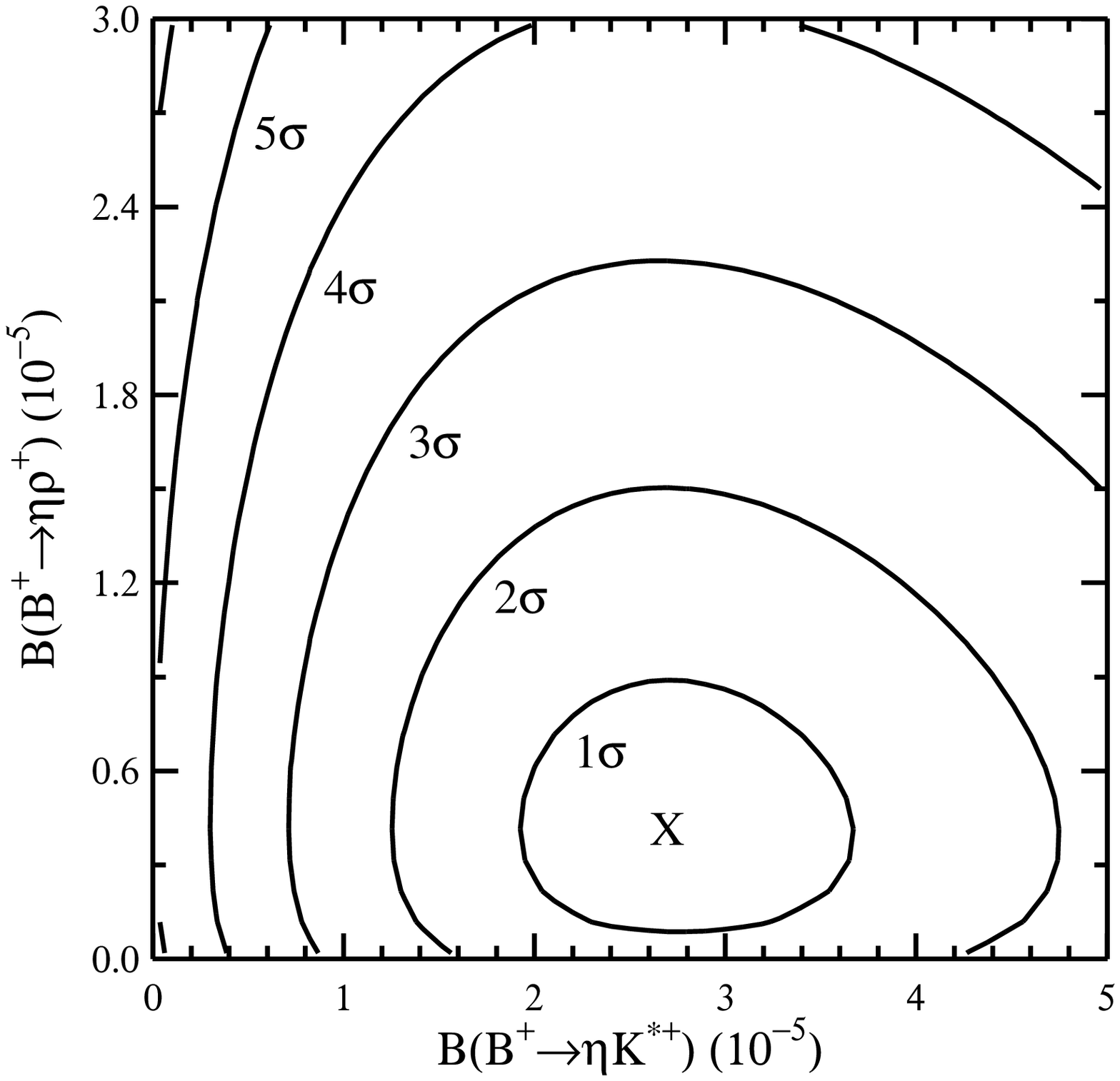}%
{60 150 530 610}{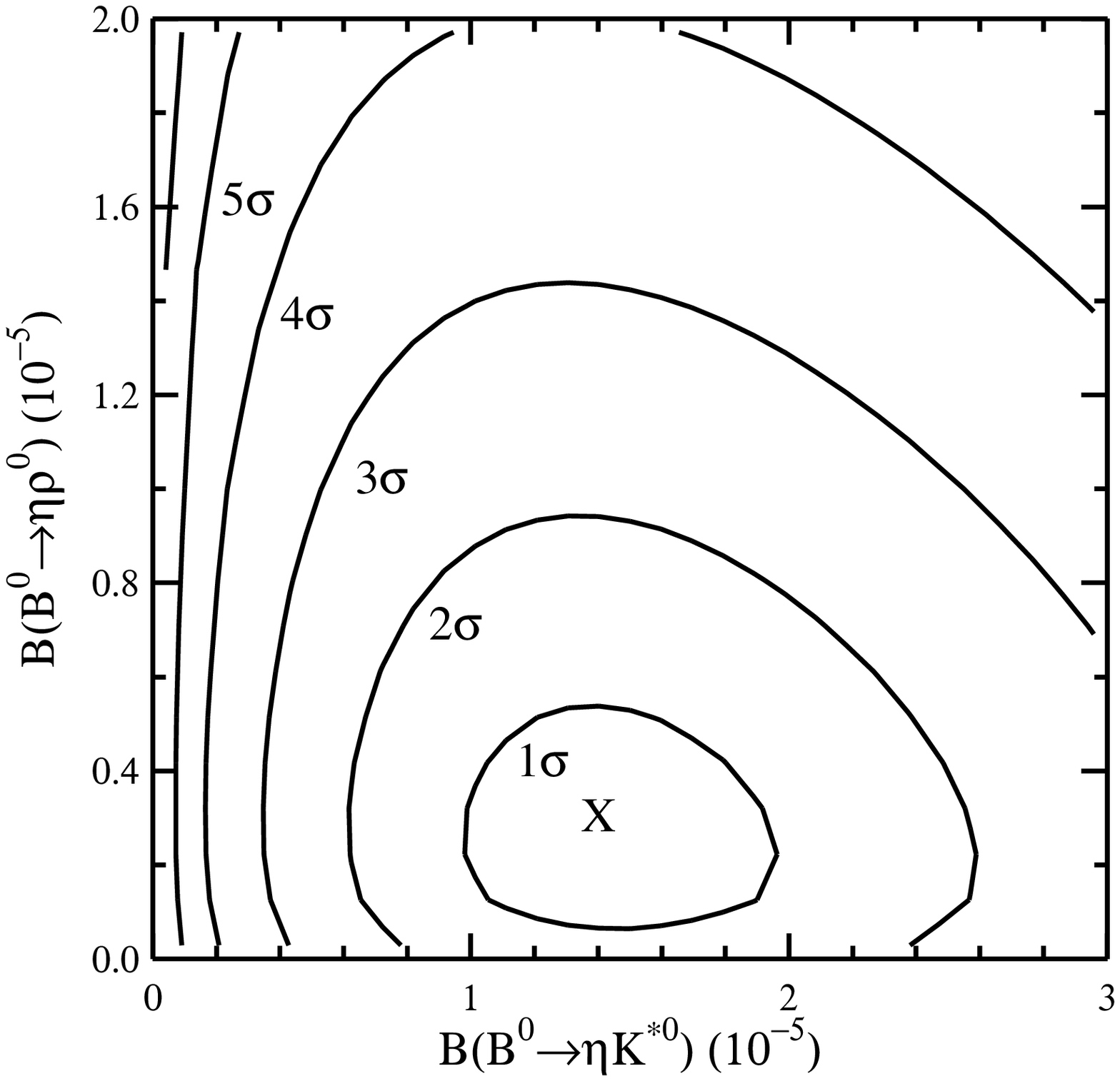}{1.0}
 \caption{\label{fig:etaCont}%
Likelihood function contours for 
(a) $B^+\ra\eta\pi^0h^+$; (b) $B^0\ra\eta\pi^-h^+$. 
 }  
\end{figure}

\begin{figure}
  \psfiletwoBB{50 140 540 610}{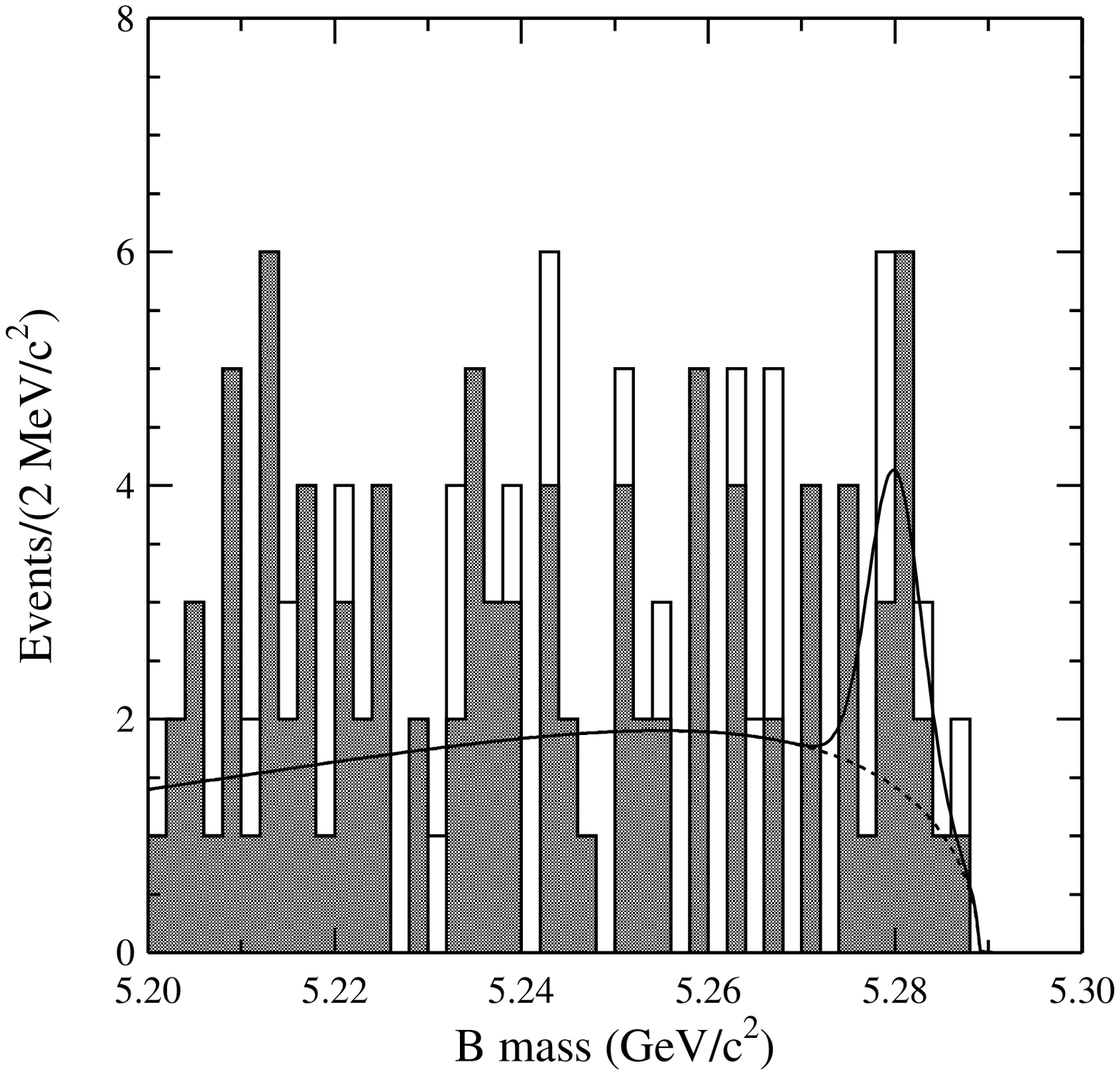}%
  {50 140 540 610}{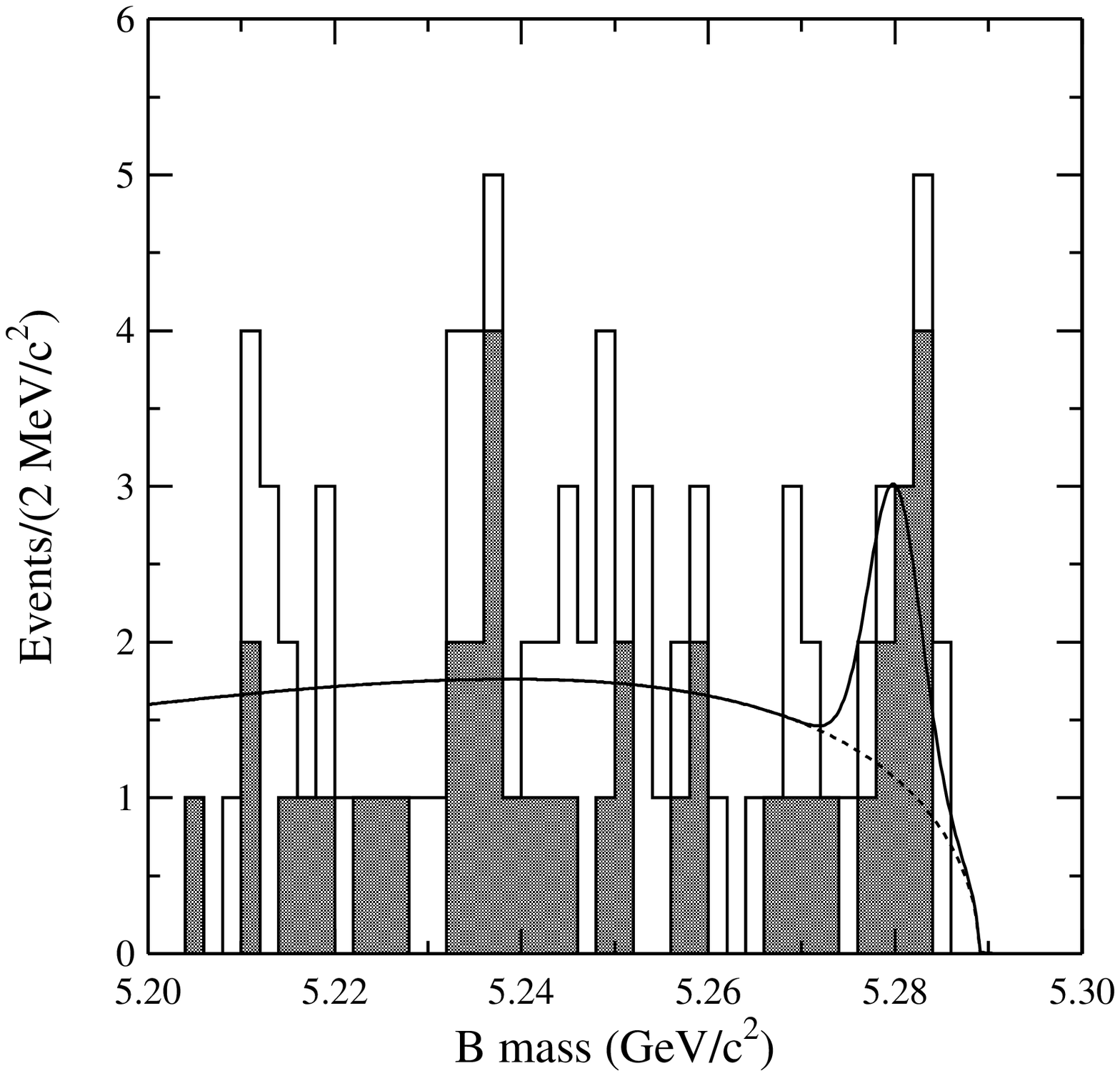}{1.0}
  \caption{\label{fig:etaKstMb}%
Projections onto the variable \mb.  Overlaid on each plot as smooth
curves are the best fit functions (solid) and background components
(dashed), calculated with the variables not shown restricted to the
neighborhood 
of expected signal.  The histograms show (a) $B^+\ra\eta
K^{*+}$; (b) \etaKstz.  The unshaded histograms represent the $\hel>0$
$K^*$ samples and the shaded is the rest.
 }
\end{figure}

The positive signals we find in both charge states of \etaKst\ are first
observations: $\BetaKstp = \RetaKstp$ and $\BetaKstz = \RetaKstz$.  (The
first error quoted is statistical, the second systematic.)  The
significance is about 5 standard deviations for both, as can be seen in 
the likelihood functions from the
fits shown in Fig.\ \ref{fig:etaCont}.  For the \etaKst\ decays we show
also in Fig.\ \ref{fig:etaKstMb}\ the projections of event distributions
onto the \mb\ axis.  The signals appear as peaks at the $B$ meson mass
of $5.28\ \GeVcsq$ in these plots, as well as in \DE\ and the $K^*$
mass.

\begin{figure}[htbp]
\psfiletwoBB{60 150 530 610}{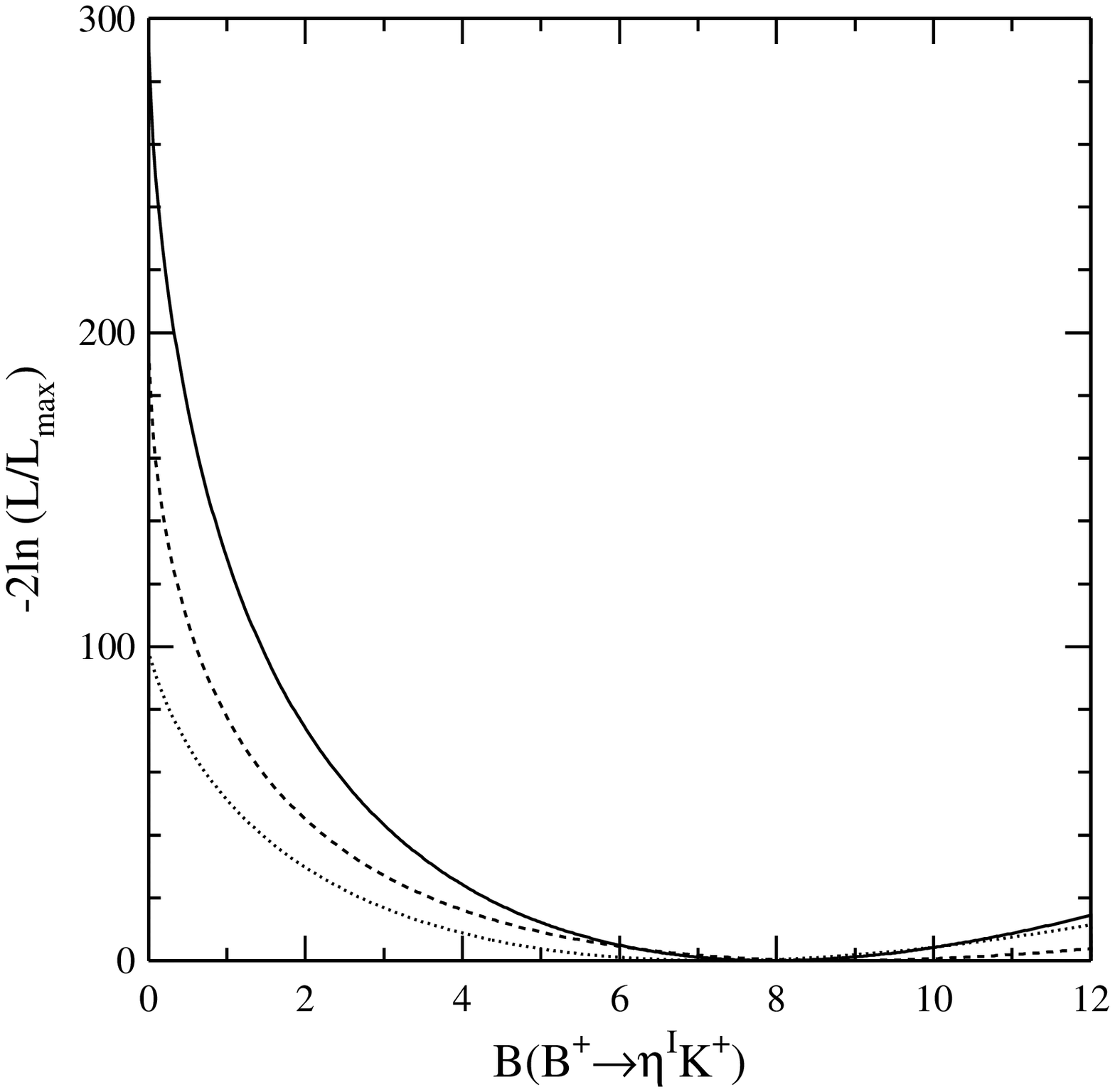}%
{60 150 530 610}{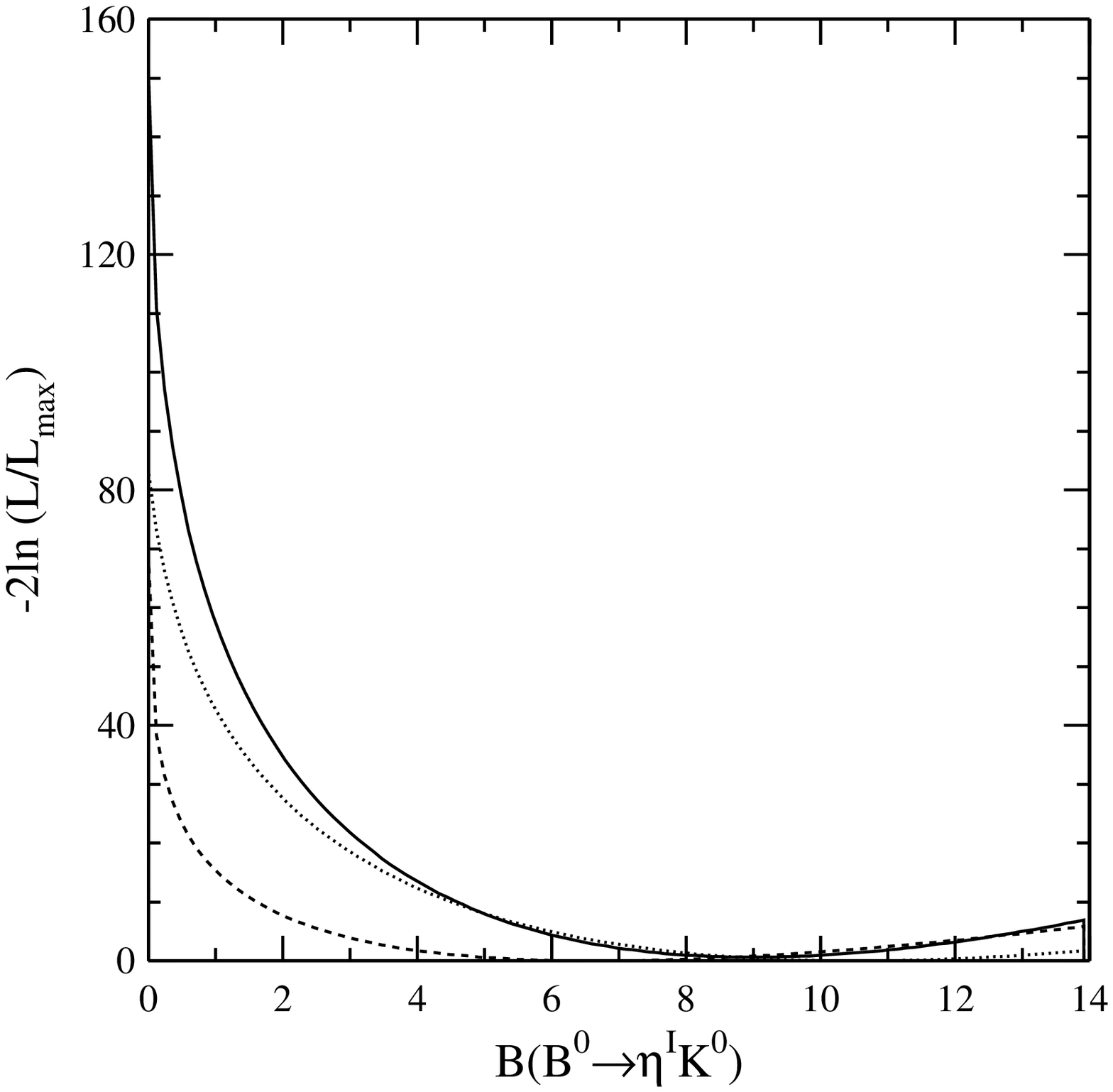}{1.0}
 \caption{\label{fig:etapCont}%
The function $-2\ln{{\cal L}/{\cal L}_{\rm max}}=\chi^2-\chi^2_{\rm min}$ for
(a) $B^+\ra\eta^\prime h^+$; (b) \etapKz.  The dashed curve is for \etaprd\
and the dotted curve for \etaprrg.
 }  
\end{figure}

\begin{figure}
  \psfiletwoBB{50 140 540 610}{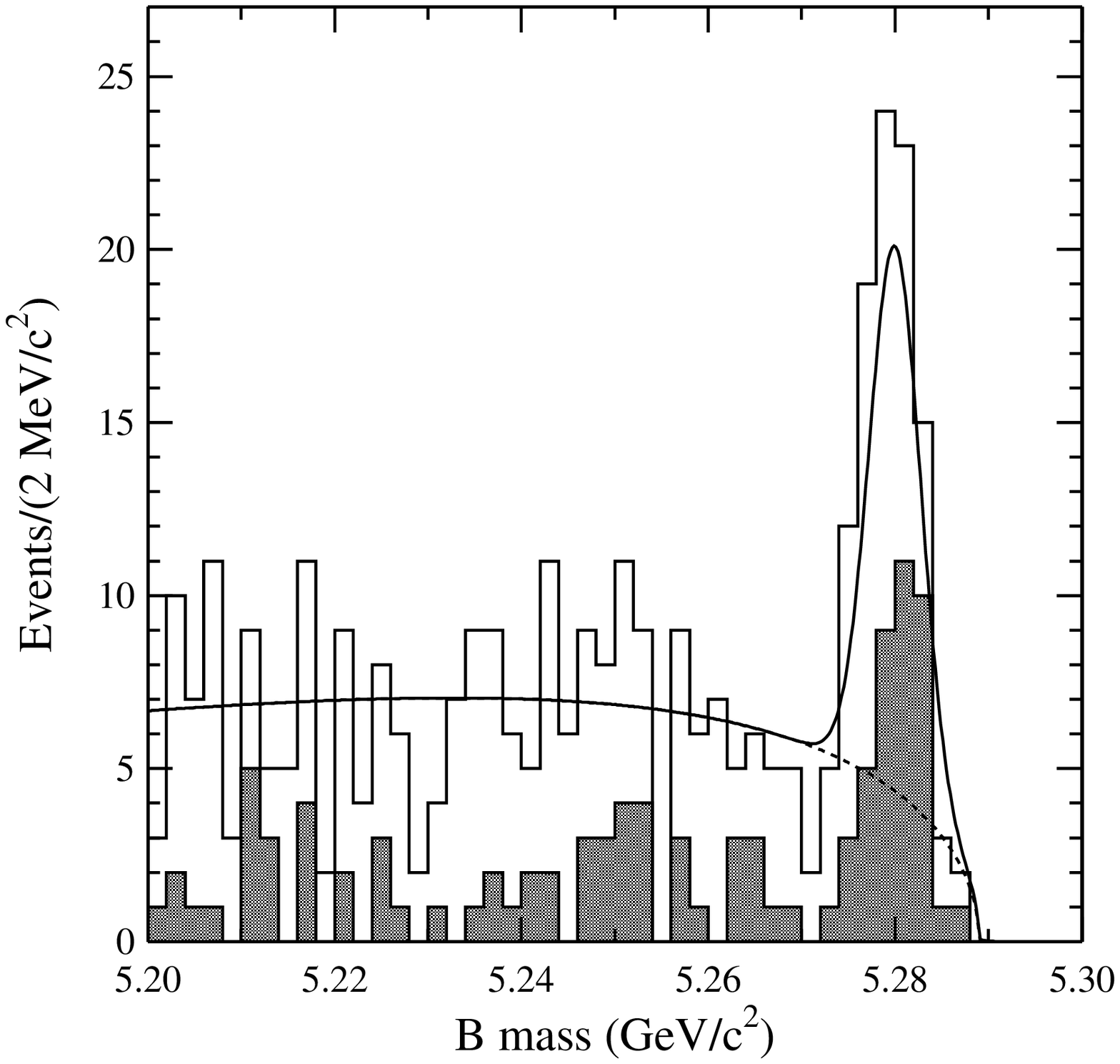}%
  {50 140 540 610}{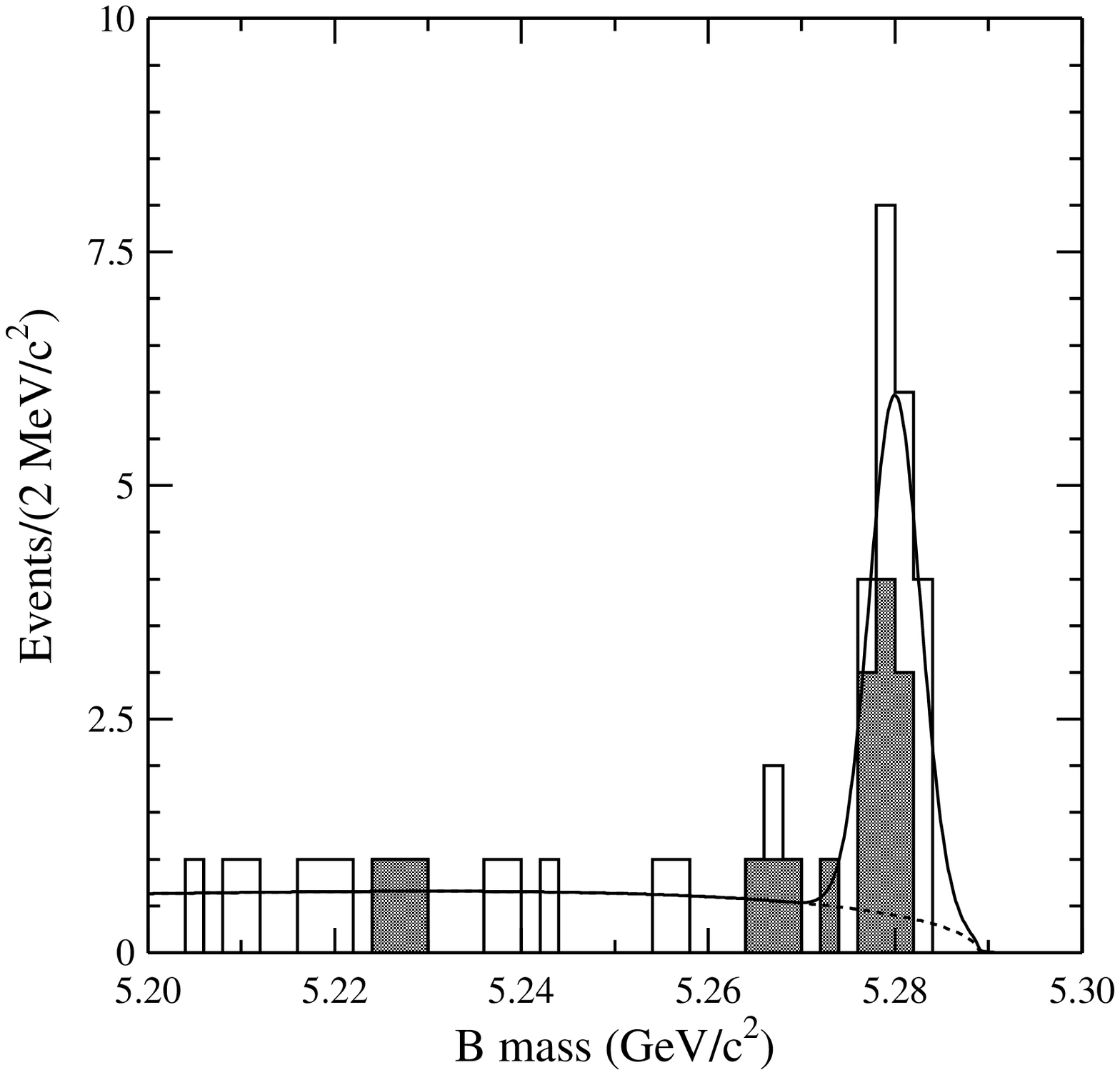}{1.0}
  \caption{\label{fig:etaphMbDe}%
Projections onto the variable \mb.  Overlaid on each plot as smooth
curves are the best fit functions (solid) and background components
(dashed), calculated with the variables not shown restricted to the
neighborhood 
of expected signal.  The histograms show (a) $B^+\ra\eta^\prime
h^+$ with $\eta^\prime\ra \eta \pi\pi\ (\eta\ra3\pi$, dark shaded), 
$\eta^\prime\ra \eta \pi\pi\ (\eta\ra\gamma\gamma$, light shaded),
and $\eta^\prime\ra \rho\gamma$ (open); (b) \etapKz\ with
$\eta^\prime\ra \eta \pi\pi$ (shaded) and $\eta^\prime\ra
\rho\gamma$ (open). 
 }
\end{figure}

With the full CLEO data sample we also improve our previous measurements
\cite{cleoEtaPRL,cleoVanc}\ of \etapK.  The new branching fractions are
$\BetapKp = \RetapKp$ and $\BetapKz = \RetapKz$.  The
likelihood functions from the fits for $B\ra \eta^\prime h^+$ and
$B^0\ra\eta^\prime K^0$ are shown in Fig.\ \ref{fig:etapCont}.  For
these modes we show also in Fig.\ \ref{fig:etaphMbDe}\ the projections
of event distributions onto the \mb\
axis.

The observed branching fractions for \etapK\ and \etaKst, in combination
with the 
upper limits for the other modes in Table \ref{combtab} and with recent
measurements of $B\ra K\pi$, $\pi\pi$ \cite{CLEObkpiNew}, and
$B\ra\pi\rho$ \cite{cleoDPF},
provide important constraints on the theoretical picture for these
charmless hadronic decays.  

The effective Hamiltonian calculations \cite{aliGreub} commonly used to
account for the charmless hadronic $B$ decays contain uncertainties in form
factors \cite{datta,kagan}, light quark masses
\cite{kagan}, and the QCD scale.  They generally employ spectator and
factorization \cite{BSW} approximations, with unknown color octet terms
arising from the expansion parameterized by an effective number of
colors.

Two approaches, with many
variations, to understanding these decay rates have been proposed.  
A large
ratio of $(\etapK,\eta K^*)$ to $(\etaK,\eta^\prime K^*)$, consistent with our
measurements, was 
predicted \cite{lipkin} in terms of
interference of the two penguin diagrams in Fig.\
\ref{fig:diagrams}(a) and (b), constructive for $\etapr K$ and $\eta K^*$
and destructive for $\eta K$ and $\etapr K^*$.  The argument is a bit
more complicated for $\eta K^*$ than $\etapr K$, and there are probably
cancellations that offset the enhancement.  With independently estimated
form factors the calculations tend to predict smaller rates than we
observe.

The second class of ideas, which address particularly the large $\etapr
K$ rate, take the $\eta^\prime$ to be a special, poorly
understood object with an anomaly contribution (Fig.\
\ref{fig:diagrams} (d)) in constructive interference with the penguins
\cite{etaCP,soni,hairpin}.  This is related to 
speculations regarding the gluon \cite{soni} or $c\bar c$ content
\cite{cheng,zhitnitsky} of the $\eta^\prime$.  These are not equivalent;
the anomaly term is a specific contribution which has been calculated
(and shown to be insufficient to explain the data).

In combination, our $\eta K^*$ and $\etapr K$ observations suggest
that perhaps the penguin interference mechanism is more important than
current estimates give, with unexpectedly large form factors or some other
mechanism favoring the penguin terms for both modes.  The models that
focus entirely on enhancement mechanisms for the \etapr\ appear to be
insufficient to explain all of the data.

We thank George Hou and Hai-Yang Cheng for many useful discussions.
We gratefully acknowledge the effort of the CESR staff in providing us with
excellent luminosity and running conditions.
J.R. Patterson and I.P.J. Shipsey thank the NYI program of the NSF, 
M. Selen thanks the PFF program of the NSF, 
M. Selen and H. Yamamoto thank the OJI program of DOE, 
J.R. Patterson, K. Honscheid, M. Selen and V. Sharma 
thank the A.P. Sloan Foundation, 
M. Selen and V. Sharma thank the Research Corporation, 
F. Blanc thanks the Swiss National Science Foundation, 
and H. Schwarthoff and E. von Toerne thank 
the Alexander von Humboldt Stiftung for support.  
This work was supported by the National Science Foundation, the
U.S. Department of Energy, and the Natural Sciences and Engineering Research 
Council of Canada.



\begin{thebibliography}{99}
\bibitem{chgconj} 
Inclusion of the charge conjugate states is implied throughout this paper.
\end{thebibliography}
\end{document}